# On the local structure of the Euler-Lagrange mapping of the calculus of variations

Demeter Krupka

**1. Introduction.** The purpose of this paper is to announce some new results on the structure of the higher order Euler-Lagrange mapping of the multiple-integral variational calculus on fibered manifolds, namely a description of its kernel and its image, and an explicit characterization of the conditions under which a system of partial differential equations (of arbitrary order) is a system of the Euler-Lagrange equations. The presented theorems complete the results obtained recently by Tonti [18], Takens [17], Kupershmidt [14], and the author [9-11, 13], and are very closely related to the research of Tulczyjew [21], Lawruk and Tulczyjew [15], and Haine [7].

A specific feature of these new results is that they are sufficiently general and very explicit. For variational problems of simple integrals, they agree with the well-known classical results (see [8] and the references therein).

Our considerations are carried out within the standard differential-geometric theory of the calculus of variations (see e.g. [2-6], [9], [20] ). The analysis of the local structure of the Euler-Lagrange mapping is based on the concept of a *Lepagean form*, introduced in [12], and its important example - the *generalized Poincare-Cartan equivalent* of a higher order Lagrangian [1], [10], [16].

**2. The Euler-Lagrange mapping.** Throughout this paper, $\pi: Y \to X$, denotes a *fibered manifold* (=surjective submersion) with orientable, $n$-dimensional base $X$ Here $j^r Y$ denotes the *r-jet prolongation* of this fibered manifold, and $\pi_r: j^r Y \to X$ and $\pi_{r,s}: j^r Y \to j^s Y$, $0 \leq s \leq r$, are the natural projections of jets. If $x^i$, $y^\sigma$ are fiber coordinates on $Y$, then the natural coordinates on $j^r Y$ relative to $x^i, y^\sigma$ are denoted by $x^i$, $y^\sigma$, $y^\sigma_{j_1}$, …, $y^\sigma_{j_1 j_2 \ldots j_r}$. Here $\Omega^p(j^r Y)$, or $\Omega^p_X(j^r Y)$, or $\Omega^p_Y(j^r Y)$ denotes the module of $p$-forms ($\pi_r$-horizontal $p$-forms, or $\pi_{r,0}$-horizontal $p$-forms) on $j^r Y$, respectively.

Recall that a *Lagrangian of order $r$* for $\pi$ is an element of $\Omega^n_X(j^r Y)$. Let $\lambda$ be such a Lagrangian. Locally

$$\lambda = \mathscr{L} \omega_0,$$

where

$$\omega_0 = dx^1 \wedge dx^2 \wedge \ldots \wedge dx^n.$$

The *Euler-Lagrange form* relative to $\lambda$ is a globally well-defined $(n+1)$-form $\mathscr{E}_\lambda$ on $j^{2r} Y$ defined by the following formulas:

$$\mathscr{E}_\lambda = \mathscr{E}_\sigma(\mathscr{L}) \omega^\sigma \wedge \omega_0,$$

$$\mathscr{E}_\sigma(\mathscr{L}) = \sum_{k=0}^{r} (-1)^k d_{j_1} \ldots d_{j_k} \frac{\partial \mathscr{L}}{\partial y^\sigma_{j_1 \ldots j_k}},$$

$$\omega^\sigma = dy^\sigma - y^\sigma_j dx^j.$$

$\mathscr{E}_\lambda$ is a $\pi_{2r,0}$-horizontal $(n+1)$-form. The mapping $\lambda \to \mathscr{E}_\lambda$ of $\Omega^n_X(j^r Y)$ into $\Omega^{n+1}_Y(j^{2r} Y)$ is called the *Euler-Lagrange mapping*. Obviously, the Euler-Lagrange mapping is **R**-linear.

**3. Lepagean forms.** In this section the Lie derivative relative to a vector field $\xi$ is denoted by $\partial_\xi$, the contraction (= inner product) of a form by $\xi$ is denoted by $i_\xi$.

The structure of a fibered manifold on $Y$ allows us to define the *horizontalization* of forms on $j^r Y$, $h: \Omega^p(j^r Y) \to \Omega^p_X(j^{r+1} Y)$, as follows: $h$ is a unique exterior-product preserving mapping such that, locally, for any function $f$ on $j^r Y$ and any integers $i$, $j_1$, …, $j_r$, $\sigma$,

$$h(f) = f \circ \pi_{r+1,r}, \quad h(dx^i) = dx^i,$$

$$h(dy^\sigma) = y^\sigma_k dx^k, \quad h(dy^\sigma_{j_1}) = y^\sigma_{j_1 k} dx^k, \quad \ldots, h(dy^\sigma_{j_1 \ldots j_r}) = y^\sigma_{j_1 \ldots j_r k} dx^k.$$

A form $\rho \in \Omega^p(j^r Y)$ is called (a) *horizontal*, if $h(\rho) = \rho$, (b) contact, or 1-*contact*, if $h(\rho) = 0$, and (c) $k$-*contact*, where $k \geq 2$, if for each $\pi_r$-vertical vector field $\xi$ on $j^r Y$, $i_\xi \rho$ is $(k-1)$-contact.

Let $\rho \in \Omega^n(j^r Y)$, let $\Xi$ be a $\pi$-vertical vector field on $Y$, $j^r \Xi$ its *r-jet prolongation* which is a vector field on $j^r Y$. We put the following question: What conditions should be satisfied by $\rho$ in order that the well-known formula

$$\partial_{j^r \Xi} \rho = i_{j^r \Xi} d\rho + d i_{j^r \Xi} \rho,$$

or the formula

$$h(\partial_{j^r \Xi} \rho) = h(i_{j^r \Xi} d\rho) + h(d i_{j^r \Xi} \rho)$$

be precisely the so called *infinitesimal first variation formula* for the Lagrangian $\lambda = h(\rho)$? Accordingly, we define: A form $\rho \in \Omega^n(j^r Y)$ is said to be *Lepagean*, if for each $\pi_{r,0}$-projectable, $\pi_r$-vertical vector field $\xi$ on $j^r Y$, $h(i_\xi d\rho)$ depends on the $\pi_{r,0}$-projection of $\xi$ only. Let $\lambda \in \Omega_X^n(j^r Y)$ be a Lagrangian. A form $\rho \in \Omega^n(j^s Y)$ is said to be a *Lepagean equivalent* of $\lambda$, if it is Lepagean, and $h(\rho) = \lambda$.

**Theorem 1.** *To each $\lambda \in \Omega_X^n(j^r Y)$ there exists a Lepagean equivalent of $\lambda$.*

The proof of Theorem 1 consists in checking that the following form is a Lepagean equivalent of $\lambda$ (we use the notation introduced before):

$$\Theta_\lambda = \mathscr{L}\omega_0 + \left(\sum_{i=1}^n \sum_{k=0}^{r-1} f_\sigma^{ij_1\ldots j_k} \omega_{j_1\ldots j_k}^\sigma \right) \wedge \omega_0,$$

where

$$f_\sigma^{j_1\ldots j_{r+1}} = 0,$$

$$f_\sigma^{j_1\ldots j_k} = \frac{\partial \mathscr{L}}{\partial y_{j_1\ldots j_k}^\sigma} - d_i f_\sigma^{ij_1\ldots j_k}, \quad k = 1, 2, \ldots, r,$$

$$\omega_{j_1\ldots j_k}^\sigma = dy_{j_1\ldots j_k}^\sigma - y_{j_1\ldots j_k l}^\sigma dx^l,$$

$$\omega_i = dx^1 \wedge \ldots \wedge dx^{i-1} \wedge dx^{i+1} \wedge \ldots \wedge dx^n.$$

The form $\Theta_\lambda$ is globally well-defined. We call it the *generalized Poincare-Cartan equivalent* of $\lambda$.

**4. The kernel of the Euler-Lagrange mapping.** The problem as to what are the Lagrangians whose Euler-Lagrange form vanishes is answered by the following

**Theorem 2.** *Let $\lambda \in \Omega_X^n(j^r Y)$. The following three conditions are equivalent:*
(1) $\mathscr{E}_\lambda = 0$.
(2) *Locally, there exists an $(n-1)$-form $\eta$ on a space $j^s Y$ such that $\lambda = h(d\eta)$.*
(3) *Locally, there exists a Lepagean equivalent $\rho$ of $\lambda$ such that $d\rho = 0$.*

**5. Naturality of the Euler-Lagrange mapping.** The following theorems express the *naturality* of the assignement $\lambda \to \Theta_\lambda$ and the Euler-Lagrange mapping $\lambda \to \mathscr{E}_\lambda$ with respect to isomorphisms of underlying fibered manifolds. In these theorems, $j^r \alpha$ denotes the $r$-*jet prolongation* of an isomorphism $\alpha$ of $\pi$.

**Theorem 3.** *The mapping $\Omega_X^n(j^r Y) \ni \lambda \to \Theta_\lambda \in \Omega^n(j^{2r-1} Y)$ is natural. In other words, for each isomorphism $\alpha$ of $\pi$,*

$$(j^{2r-1}\alpha)^* \Theta_\lambda = \Theta_{j^r\alpha *\lambda}.$$

**Theorem 4.** *The mapping $\Omega_X^n(j^r Y) \ni \lambda \to \mathscr{E}_\lambda \in \Omega_Y^{n+1}(j^{2r} Y)$ is natural. In other words, for each isomorphism $\alpha$ of $\pi$,*

$$(j^{2r}\alpha)^* \mathscr{E}_\lambda = \mathscr{E}_{j^r\alpha *\lambda}.$$

Note the following corollary of Theorem 4 and Theorem 2. Let $\alpha$ be a *generalized invariant transformation* of $\lambda$, i.e.,

$$(j^{2r}\alpha)^*\mathcal{E}_\lambda = \mathcal{E}_\lambda.$$

Then locally,

$$\lambda - j^r\alpha^*\lambda = h(d\eta)$$

for some $(n-1)$-form $\eta$. The infinitesimal version of this formula is known as the *Noether-Bessel-Hagen equation* [19], [20].

**6. The image of the Euler-Lagrange mapping and the inverse problem of the calculus of variations.** We state the following definitions. Consider any form $\varepsilon \in \Omega_X^{n+1}(j^sY)$ such that for each $\pi_s$-vertical vector field $\xi$ on $j^sY$, $i_\xi\varepsilon \in \Omega_X^n(j^sY)$. Obviously, each Euler-Lagrange form has this property. We say that $\varepsilon$ is *variational* (resp. *weakly variational*), if there exist an integer $r$ and a Lagrangian $\lambda \in \Omega_X^n(j^rY)$ such that $\varepsilon = \mathcal{E}_\lambda$ (resp. such that for each section $\gamma$ of $\pi$ satisfying $\varepsilon \circ j^s\gamma = 0$, the condition $\mathcal{E}_\lambda \circ j^{2r}\gamma = 0$ holds). Analogously, the notions of *local variationality* and *weak local variationality* are introduced.

**Theorem 5.** *Let $\varepsilon \in \Omega_Y^{n+1}(j^sY)$ be a form such that for each $\pi_s$-vertical vector field $\xi$ on $j^sY$, $i_\xi \in \Omega_X^n(j^sY)$. Then $\varepsilon$ is locally variational if and only if to each point of $j^sY$ there exist a neighborhood W of this point, an integer $r \geq s$, and a 2-contact $(n+1)$-form $\eta_W$ defined on $\pi_{r,s}^{-1}(W)$ such that*

$$d(\varepsilon + \eta_W) = 0.$$

Let now $\varepsilon$ be locally variational. The question how to find a Lagrangian for $\varepsilon$ is answered by the following

**Theorem 6.** *Assume that $\varepsilon$ is locally variational, and write, locally,*

$$\varepsilon = \varepsilon_\sigma\, \omega^\sigma \wedge \omega_0.$$

*Then the formulas*

$$\lambda = \mathcal{L}\omega_0, \quad \mathcal{L} = y^\sigma \int_0^1 \varepsilon_\sigma(x^i, ty^\sigma, \ldots, ty^\sigma_{j_1\ldots j_s})\,dt$$

*define, locally, a Lagrangian $\lambda$ for which $\mathcal{E}_\lambda = \varepsilon$.*

The Lagrangian $\lambda$ of Theorem 6 is called the *Tonti Lagrangian*. In the paper [18] one can find many examples of these Lagrangians, with the integration performed explicitly.

It follows from Theorem 6 that if $\pi: Y \to X$ is a *vector* bundle and $\varepsilon$ is locally variational, then $\varepsilon$ is (globally) variational.

**7. Variationality of systems of differential equations.** Consider a system

$$G_i(t, q_k, \dot{q}_k, \ddot{q}_k) = 0,$$

of ordinary differential equations for functions $t \to q_k(t)$. We ask under what conditions does this system coincide with some Euler-Lagrange equations. It is well-known that is so if and only if the following *Helmholtz conditions* hold [8]:

$$\frac{\partial G_i}{\partial q_k} - \frac{\partial G_k}{\partial q_i} = \frac{1}{2}\frac{d}{dt}\left(\frac{\partial G_i}{\partial \dot{q}_k} - \frac{\partial G_k}{\partial \dot{q}_i}\right),$$

$$\frac{\partial G_i}{\partial \dot{q}_k} + \frac{\partial G_k}{\partial \dot{q}_i} = \frac{d}{dt}\left(\frac{\partial G_i}{\partial \ddot{q}_k} + \frac{\partial G_k}{\partial \ddot{q}_i}\right),$$

$$\frac{\partial G_i}{\partial \ddot{q}_k} - \frac{\partial G_k}{\partial \ddot{q}_i} = 0.$$

We shall now generalize the Helmholtz conditions to partial differential equations of any order.

With our standard notation, consider a system

$$(*) \qquad \varepsilon_\sigma(x^i, f^\nu(x^i), D_{j_1}f^\nu(x^i), \ldots, D_{j_1}\ldots D_{j_r}f^\nu(x^i)) = 0$$

of partial differential equations, where $1 \le i \le n$, $1 \le \sigma, \nu \le m$, and $r$ is an arbitrary integer. We shall say that this system is *variational*, if the form $\varepsilon = \varepsilon_\sigma \omega^\sigma \wedge \omega_0$ is variational.

**Theorem 7.** *The system* $(*)$ *is variational if and only if*

$$\frac{\partial \varepsilon_\sigma}{\partial y^\nu} - \frac{\partial \varepsilon_\nu}{\partial y^\sigma} - \sum_{k=1}^{r}(-1)^k d_{j_1}\ldots d_{j_k}\frac{\partial \varepsilon_\nu}{\partial y^\sigma_{j_1\ldots j_k}} = 0,$$

$$\frac{\partial \varepsilon_\sigma}{\partial y^\nu_{j_1\ldots j_l}} - (-1)^l\frac{\partial \varepsilon_\nu}{\partial y^\sigma_{j_1\ldots j_l}} - \sum_{k=l+1}^{r}(-1)^k\binom{k}{l}d_{j_{l+1}\ldots j_k}\frac{\partial \varepsilon_\nu}{\partial y^\sigma_{j_1\ldots j_k}} = 0, \quad 1 \le l \le r.$$

We shall say that the system $(*)$ is *weakly variational*, if the form $\varepsilon = \varepsilon_\sigma \omega^\sigma \wedge \omega_0$ is weakly variational. Theorem 7 is the key to the weak variationality of the system $(*)$ or, which is the same, to the problem of extention of the range of applicability of the Lagrange formalism to systems of equations, which are not necessarily variational. Obviously, in order that $(*)$ be weakly variational it is enough that there exist functions $A_\sigma^\rho$ such that the form $\varepsilon' = \varepsilon'_\sigma \omega^\sigma \wedge \omega_0$, where $\varepsilon' = A_\sigma^\rho \varepsilon_\rho$, is variational.

**Acknowledgement.** The author is gradeful to Dr. M. Lenc from the Institute of Scientific Instruments of the CSAV for having drawn his attention to the inverse problem of the calculus of variations, and for many critical and stimulating discussions.

**Author's address:** Department of Mathematics, Faculty of Science, Purkyne University, 662 95 Brno, CZECHOSLOVAKIA